\documentclass[usenatbib,usegraphicx]{mn2e}

\usepackage{newtxtext,newtxmath}
\usepackage{txfonts}

\usepackage[T1]{fontenc}
\usepackage{ae,aecompl}

\usepackage{graphicx}	
\usepackage{amsmath}	
\usepackage{amssymb}	
\usepackage{color}

\title[Absolute Properties of RU Cnc]{
Absolute properties of RU Cnc revisited: {\LARGE{An active RS CVn-type eclipsing binary with a red giant branch and a main sequence components}}}

\author[\c{C}okluk et al.]
{
    K. A. \c{C}okluk,$^{1}$ 
    {D. Ko\c{c}ak,$^{1}$\thanks{Corresponding author: Dolunay  Ko\c{c}ak 
    	E-mail: dolunay.kocak@gmail.com}} 
    T. {\.I}\c{c}li,$^{1}$ 
    S. Karak\"{o}se,$^{1}$
    S. \"{U}st\"{u}nda\u{g},$^{1}$
and K. Yakut$^{1}$   \\
   {\normalsize $^{1}$Department of Astronomy and Space Sciences, University of Ege, 35100, {\.I}zmir, Turkey}}
\date{Accepted XXX. Received YYY; in original form ZZZ}

\pubyear{2019}

\begin{document}
\label{firstpage}
\pagerange{\pageref{firstpage}--\pageref{lastpage}}
\maketitle
\begin{abstract}
We present observations and analysis of an RS CVn-type double-lined eclipsing binary system, RU Cnc.
The system has been observed for over a century.
The high-quality long-cadence \emph{Kepler} K2 C5 and C18, newly obtained observations,
and two radial velocity curves were combined and analyzed simultaneously assuming multi-spot model.
The masses, radii and luminosities of the component stars are precisely obtained as $M_\textrm{c} = 1.386\pm0.044\, M_{_\odot}$, $M_\textrm{h} = 1.437 \pm 0.046\, M{_\odot}$, $R_\textrm{h} = 2.39\pm 0.07\, R{_\odot}$, $R_\textrm{c} = 5.02 \pm 0.08\, R{_\odot}$, $L_\textrm{h} = 11.4\pm 1.2\, L{_\odot}$, $L_\textrm{c} = 12.0 \pm 1.0\, L{_\odot}$ and with a separation of $\textrm{a} = 27.914 \pm 0.016\, R{_\odot}$. The distance of the system is determined to be $380\pm 57\,$ pc which is consistent with the Gaia DR2 result. Long-term detailed period variation analysis of the system indicate a period decrease of $7.9\times10^{-7}$ days per year. The results suggest the cooler component to be on the red giant branch (RGB) and the hotter one to be still on the main sequence.
\end{abstract}




\section{INTRODUCTION}
The system RU Cnc (HIP 42303, BD +24$^\circ$1959,  EPIC 212173112) is an RS CVn-type eclipsing variable system with an orbital period of 10.2 days. The light variation of the system shows total eclipse. The spectral type of the components are G9V and F9V (Wyse 1934)
with a strong Ca II H \& K spectral lines (Struve 1945).
Although good radial velocity curves exist, the full light curve of RU Cnc could not be obtained until recently,
possibly because of the relatively long orbital period of the system.

The system RU Cnc first observed by van Biesbroeck and Casteels (1914).
Russell and Shapley (1914) gave the period and parallax of the binary system as 10.4 days 2.6 mas,
which is very close to the current Gaia DR2 value. Fetlaar (1930) determined some orbital parameters of the RU Cnc.
Nijland (1931), Lause (1939), Sandig (1947), and Szafraniec (1957) observed the system variability.
Popper and Dumont (1977) and Scaltriti (1979) reported an anomaly at the orbital phase range 0.65-0.85.
Similarly, Cerruti-Sola et al. (1980) indicated a brightening at the orbital phase of 0.75.
Blanco et al. (1983) claimed these anomalies to be due to the reflection or ellipticity.
On the other hand, Busso et al. (1984) indicated Solar-like spot activity may cause this anomaly.

Popper (1990) obtained mass and radius of the components as 
$M_\textrm{c} = 1.47\, M{_\odot}$, $M_\textrm{h} = 1.46\, M{_\odot}$, $R_\textrm{h} = 1.9\, R{_\odot}$, $R_\textrm{c} = 4.9\, R{_\odot}$.
Eggleton and Yakut (2017) modeled 60 binary/multiple systems containing giant stars.
In their study, they calculated evolutionary models for the system RU Cnc.
The authors adopted the physical parameters of the systems from the literature as $M_\textrm{c} = 1.35\, M{_\odot}$, $M_\textrm{h} = 1.41\, M{_\odot}$, $R_\textrm{h} = 1.97\, R{_\odot}$, $R_\textrm{c} = 4.85\, R{_\odot}$, T$_{\textrm c}$=4800~K and T$_{\textrm h}$=6400~K.
It has been also mentioned the necessity to re-obtain some of the observational parameters.
In this study, we revisited this system which is important to stellar evolution.

Nowadays, there are many surveys and satellite projects to observe stars (e.g, Kepler, Gaia, TESS).
Highly sensitive photometric observations of \emph{Kepler} satellite led many stellar and
planetary discoveries (\textit{e.g.}, Rappaport et al. 2017).
In addition to very precise data continuous observations, especially for the RS CVn-type systems, provide important results.
Following four years mission, \emph{Kepler} continued its observations with the K2 mission
(Howell et al. 2014). During the K2 mission \emph{Kepler} observed the system RU Cnc in the K2 C5 and C18 campaigns.
In this study, two data sets of the \emph{Kepler} K2 observations of an active RS Cvn-type binary system RU Cnc have been studied with an Earth-based robotic telescope's data and radial velocity curves obtained at two different times (Popper, 1990 and Imbert, 2002).

\section{OBSERVATIONS}
\begin{figure}
\centering
\includegraphics[height=120mm]{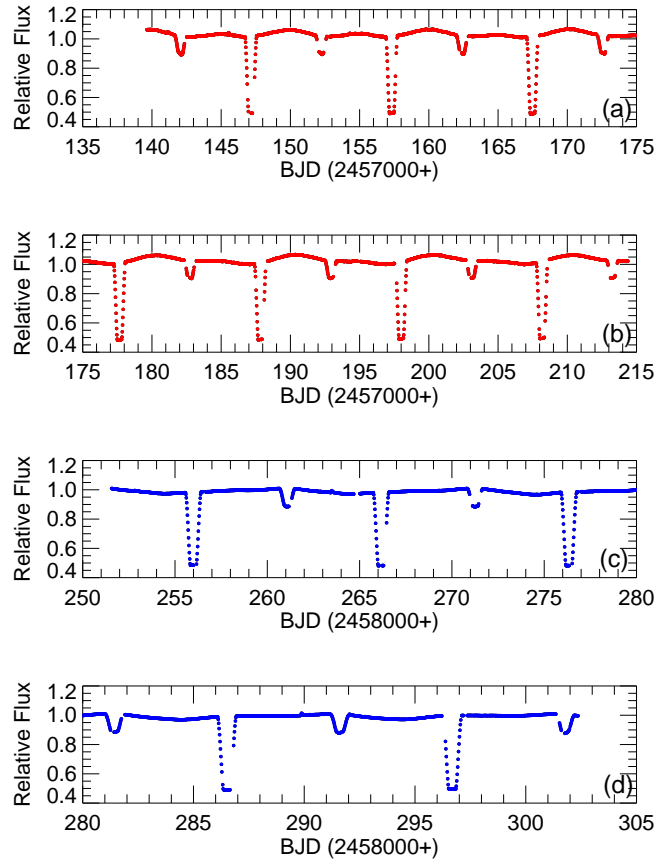}\\
\caption{Long-cadence \emph{Kepler} K2 C5 (red, a,b) and C18 (blue, c, d) observations of RU Cnc (see Section 3 for details).}
\label{Fig:rucnc_kepler}
\end{figure}

\begin{figure}
\centering
\includegraphics[height=60mm,width=90mm]{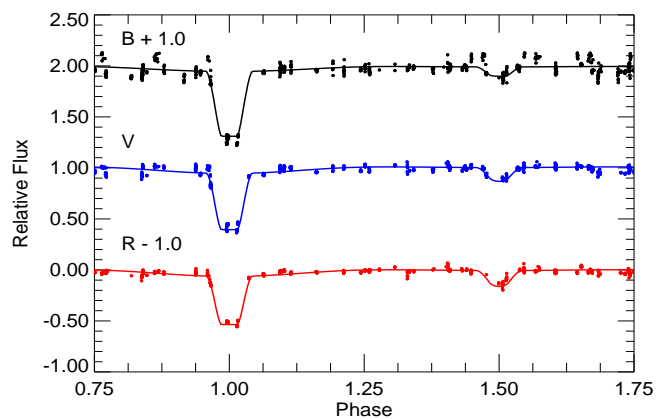}\\
\caption{TUG T60 observations of RU Cnc in BVR bands. B and R bands are shifted by value of a 1.0. See the text for the details.}\label{Fig:lct60}
\end{figure}

RU Cnc was observed by the \emph{Kepler} K2 mission on 75 days during the campaign C5 in 2015 and 36 days during the campaign C18 in 2018. Normalized flux light curve of the binary system is shown in the Figures~\ref{Fig:rucnc_kepler}a-d and the data are given Table \ref{tab:data2}.
Full data of the system with eclipses are obtained. All the Kepler K2 data corrected by using LcTools (Kipping et al. 2015).
In addition, long-term observations of RU Cnc have been performed by a 60-cm Earth-based robotic
telescope at the T\"UB\.ITAK National Observatory (TUG) in the years 2016 and 2017.
BVR filters were used to observe the systems on 67 nights. 2K$\times$2K FLI Proline CCD has been used during the observations.
Reduction of the data was made with IRAF reduction packages, AstroImageJ (Collins et al. 2017) and some Python codes developed by us.
Reductions were performed by subtracting the bias and dark dividing by the flat. Following the time correction
differential photometry have been performed. As a comparison star TYC 1942-1809-1 (C1), 2MASS 08370964+2328007 (C2), 
2MASS 08371367+2328503 (C3) and 2MASS 08375144+2329117 (C4) were used.
The Figure~\ref{Fig:lct60} shows the light variation of the system obtained at TUG.
It is apparent from the Figure~\ref{Fig:lct60} that eclipses with good phase coverage could not be obtained. 
On the other hand, these observations are useful to study the long-term stellar activity.
We will discuss photmetric variation in more detail in the next section.

We obtained 22 minimum times throughout \emph{Kepler} K2 observations. 
They are listed in Table \ref{tab:minimatimes}, together with those published in the literature. 
Using the new  \emph{Kepler} minimum times we derived the linear ephemeris given in the Eq.\ref{eq1}.
During the calculation of the orbital phases in the Figures \ref{Fig:lct60} - \ref{Fig:rucnc_lcrv} and Table \ref{tab:data} we used the following ephemeris:

\begin{equation} \label{eq1}
\textrm{HJD~Min~I} = 24~57147^\textrm{d}.1401(3)+10^\textrm{d}.172928(26).
\end{equation}

\begin{table}
\caption{Basic properties of RU Cnc from the {\it Kepler} Input Catalogue (KIC) and Simbad.}
\begin{tabular}{llll}
\hline
Kepler K2 ID              &&& EPIC 212173112                  \\
2MASS ID                  &&& J08373012+2333418          \\
Gaia ID                   &&& DR1 666399497785893120    \\
$\alpha_{2000}$           &&& 08:37:30.13               \\
$\delta_{2000}$           &&& 23:33:41.6                \\
B                         &&& 10\fm 80            \\
V                         &&& 10\fm 20            \\
2MASS J                   &&& 8\fm 618            \\
2MASS H                   &&& 8\fm 140            \\
K$_p$(\textit{Kepler})    &&& $9\fm 91$                \\
Spectral type             &&& F5 V + K1 IV             \\
\hline
\end{tabular}
\label{tab:data}
\end{table}

\begin{table}
	\begin{center}
		\scriptsize \caption{Kepler K2 (C5, C18) and TUG T60 (BVR) measurements of RU Cnc.}\label{tab:data2}
		\begin{tabular}{llll}
\hline
BJD (2400000+)	 & Phase	& Normalized Flux   & Notes \\
\hline
57139.630996	&	0.2619	&	1.033214	&	1	\\
57139.651428	&	0.2639	&	1.033458	&	1	\\
57139.671860	&	0.2659	&	1.033559	&	1	\\
57139.692292	&	0.2679	&	1.033646	&	1	\\
57139.712724	&	0.2699	&	1.033744	&	1	\\
57139.733156	&	0.2719	&	1.033801	&	1	\\
57139.753588	&	0.2739	&	1.033739	&	1	\\
57139.774020	&	0.2759	&	1.033825	&	1	\\
57139.794452	&	0.2779	&	1.034038	&	1	\\
57139.814884	&	0.2799	&	1.033900	&	1	\\
57139.835316	&	0.2819	&	1.034085	&	1	\\
57139.855749	&	0.2839	&	1.033757	&	1	\\
57139.876181	&	0.2860	&	1.033963	&	1	\\
57139.896613	&	0.2880	&	1.033953	&	1	\\
\hline
\end{tabular}
\end{center}
{Notes. See Figures \ref{Fig:rucnc_kepler} and \ref{Fig:lct60} . The phases were calculated using Eq.~\ref{eq1}. In the fourth column, 1, 2, 3, 4 and 5 denote the Kepler K2 C5, Kepler K2 C18, TUG T60 B, TUG T60 V and TUG T60 R filters, respectively. Table \ref{tab:data2} is published in its entirety in the electronic edition of the MNRAS and CDS. A portion is shown here for guidance regarding its form and content.}
\end{table}

\section{DATA ANALYSIS}

\subsection{Activity}
It has been known for a long time that RS CVn-type binary systems show strong stellar activities
(see e.g Rodono et al. (1995), Morgan \& Eggleton (1979), Eaton \& Hall (1979), Hall (1976)).
Stellar activity in these systems exist as stellar spots that are detectable photometrically and spectroscopically.
Stellar spots activity mostly depends on the rotation period of the star therefore on the binary period for synchronized systems and on the convective layer (see e.g. Yakut \& Eggleton 2005, Yakut 2006). Light curves obtained at various times show variability, particularly at the maximum light. These provide information about the variation of stellar spots over time. These maximum light variations are visible in the newly obtained multi-color light variations with the 60-cm robotic telescope and Kepler K2 light variations (see Figure \ref{Fig:rucnc_kepler}).
\par
\begin{figure}
\centering
\includegraphics[height=60mm]{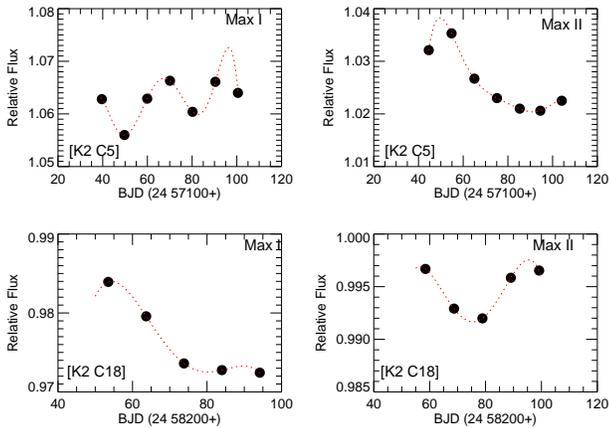}\\
\caption{Activity behavior at the Max I (Phase 0.25) and at the Max II (Phase 0.75) for light curves of an RS CVn-type system RU Cnc.}\label{Fig:activity}
\end{figure}

Since the Kepler data is much more precise, it has been obtained the relative flux for each data set. 
In the Fig.~\ref{Fig:activity} average flux values obtained for the 0.25 phase (Max~I) and for the 0.75 phase (Max~II) 
are shown versus time for the K2 C5 and C18 datasets.
Significant light variation is visible in the Fig.~\ref{Fig:activity}. 
Particularly relative flux variations during the Max~I is about a month for the C5 data and during Max~II of C18 is about 40 day.
This indicates that the stellar surface is active with stellar spots.
In the following section, light curve analysis is done taking into account the existence of stellar spots.

\subsection{Period Change Analysis}
\begin{table}
\begin{center}
\scriptsize \caption{The primary and the secondary minima times in HJD*(HJD - 2\,400\,000) for RU Cnc.}\label{tab:minimatimes}
\begin{tabular}{llllll}
\hline
HJD$*$ & Ref.& HJD$*$ & Ref. & HJD$*$ & Ref.\\
\hline
19090,164	&	1	&	22243,798	&	2	&	57142,07684	&	14	\\
19141,056	&	2	&	22416,736	&	2	&	57147,13563	&	14	\\
19151,243	&	1	&	22650,723	&	2	&	57162,41536	&	14	\\
19690,392	&	2	&	23464,572	&	2	&	57167,48478	&	14	\\
19710,735	&	2	&	24146,147	&	2	&	57172,58806	&	14	\\
19741,239	&	2	&	24949,817	&	2	&	57177,65737	&	14	\\
19771,779	&	2	&	25183,775	&	3	&	57187,82460	&	14	\\
19822,643	&	2	&	25244,841	&	2	&	57192,92599	&	14	\\
20148,161	&	2	&	25590,726	&	2	&	57198,00297	&	14	\\
20158,334	&	2	&	27106,457	&	2	&	57203,09605	&	14	\\
20168,517	&	2	&	27157,417	&	4	&	57208,17546	&	14	\\
20239,735	&	2	&	27503,218	&	5	&	57213,26712	&	14	\\
20514,381	&	2	&	28835,888	&	6	&	58255,99055	&	14	\\
20534,738	&	2	&	28937,610	&	6	&	58261,07635	&	14	\\
20585,597	&	2	&	29334,404	&	5	&	58266,16097	&	14	\\
20595,767	&	2	&	35224,470	&	7	&	58271,26396	&	14	\\
20616,122	&	2	&	36628,445	&	8	&	58276,33492	&	14	\\
20829,748	&	2	&	41775,856	&	9	&	58281,43136	&	14	\\
21185,795	&	2	&	45356,713	&	10	&	58286,50770	&	14	\\
21358,733	&	2	&	48500,200	&	11	&	58291,60874	&	14	\\
21643,592	&	2	&	52630,333	&	12	&	58296,68079	&	14	\\
21735,132	&	2	&	53769,732	&	13	&	58301,77198	&	14	\\
\hline
\end{tabular}
\end{center}
{References for Table~\ref{tab:minimatimes}. 
	1-van Biesbroeck et al.(1914),
	2-Nijland (1931);
	3-Kordylewski (1928);
	4-Sanding et al.(1947);
	5-Szafraniec (1956);
	6-Lause (1939);
	7-Szafraniec (1957);
	8-Huth (1964);
	9-Popper et al. (1977);
	10-Scaltriti (1985);
	11-Hipparcos (ESA, 1997);
	12-ASAS (Paczy{\'n}ski et al. 2006);
	13-Meyer (2008);
	14-This study}
\end{table}

The system RU Cnc has a relatively long orbital period which limits the observed times of minimum. On the other hand, long-term follow-up observation of the system has provided time-of-minima that spanned over a long time.
Erdem \& \"Ozt\"urk (2014) along with other systems performed period analysis of RU Cnc.
The authors estimated an orbital period decreases of 0.11(3) s/yr. 
In this study, we collected 66 the times of minimum obtained with visual, photographic, and new precise Kepler K2 observations.

\begin{figure}
\centering
\includegraphics[height=100mm]{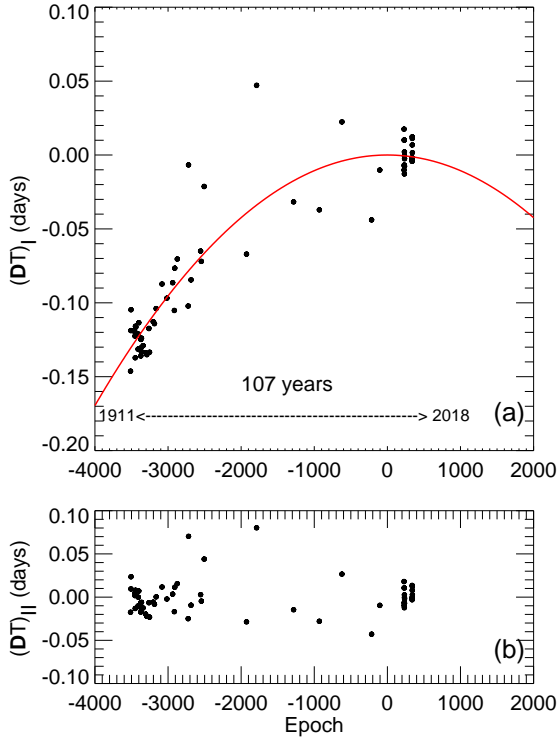}\\
\caption{(a) The (O-C)$_I$ \,($DT_I$) diagram of the times of mid-eclipses for RU Cnc constructed using the light elements given in Eq. (2) and (b) the (O–C)$_{II}$ \, ($DT_{II}$) residuals after subtraction of parabolic variation.}
\label{Fig:rucnc_oc}
\end{figure}

We calculated 22 new minimum times by using the method of Kwee \& van Woerden (1956) from the Kepler K2 data. 
They are listed in Table \ref{tab:minimatimes}, together with those published in the existing literature (Table~\ref{tab:minimatimes}).
We used Eq. \ref{eq2}, along with the minima times given in Table~\ref{tab:minimatimes} and a weighted least-squares solution, 
to estimate parameters of period change analysis. 
During the analysis we assigned weight 10 to Kepler K2 data, 5 to photoelectric, 2 to photographic (pg), and 1 to visual data.
In the Figure~\ref{Fig:rucnc_oc}, variations obtained from the period change analysis are shown.
A parabolic variation is visible in the upper panel of the Figure~\ref{Fig:rucnc_oc}.

Different assumptions are made for the orbital change analysis.
First,  for a parabolic variation, we used the least square method to analyze the data.  The analysis gave the equation:
\begin{equation}\label{eq2}
\small
\textrm{HJD~Min~I} = 24~54827^\textrm{d}.724(16) + 10^\textrm{d}.172904(6)E - 1.1(2)\times 10^{-8} E^2
\end{equation}

We estimated the period change rate ($\frac{dP}{dt}$) decrease as $7.9(1.2)\times 10^{-7}$ days  per year using the parameters given in the Eq.~(2).
On the other hand, since the system is a detached binary no mass transfer between the components can take place.
In addition, the cooler component in the system appears to be very active (Section~3.3).
This may show the existence of an important amount of mass loss from the system.
It is known that stellar activity affects the light curve and therefore the times of minima of binary systems. 
Indeed, precise and continuous Kepler data show the eclipse timing variations explicitly (\textit{e.g.}, Rappaport et al., 2013; Tran et al., 2013; Almeida et al. 2019). This might be one of the reasons for the observed such a scattering in the O-C curve of  RU Cnc.
The main reason for the period change could be also explained by the existence of a relatively
long period third body in the system with a mass loss from the system.
For estimating the period of the third body, we used Eq. (2) of Kalomeni et al. (2007). 
The data obtained for the system so far is not sufficient to make any conclusion while we conclude that 
the orbital period of a possible third body in the system should not be shorter than two century.

\subsection{Modeling Light and Radial Velocity Curves}
\par
Automatized analysis of binary stars that are observed with the \emph{Kepler} telescope
were done by Pr{\v s}a et al. (2011) and the orbital parameters of many systems have been determined.
Light curve analysis of an individual binary system naturally provides more accurate results than those.
Especially if the radial velocity curve of a binary system exists obtained physical parameters are very accurate (e.g., Yakut et al. 2015, Rappaport et al. 2016).
\par
Spectroscopic study of the system RU Cnc has been performed by Popper (1990) and Imbert (2002).
None of the photometric studies of the system exist in the literature contain full light curve.
The new light curve of the system obtained in this study does not cover the eclipses well.
This made us use only the \emph{Kepler} K2 data and radial velocity curves during the analysis.
We used Phoebe (Pr{\v s}a and Zwitter, 2005; Wilson \& Devinney, 1971, Wilson, 1979) program for light and radial velocity curves modeling.
For the simultaneous solutions of the light curve and radial velocity the curve-dependent weights were assigned as described by Wilson (1979).

\emph{Kepler} K2 data and two radial velocity curves are modeled simultaneously.
The temperature of the cooler star T$_{\rm c}$, from Eggleton and Yakut (2017),
the limb darkening coefficients $x_{\rm c}$ and $x_{\rm h}$, from Claret (2018), albedos
A$_{\rm c}$ and A$_{\rm h}$, from Rucinski (1969), the values of the gravity-darkening coefficients
g$_{\rm c}$ and g$_{\rm h}$, from Lucy (1967) were taken as fixed parameters.
During the simultaneous solution; the semi-major axis of the relative orbit (a),
binary center-of-mass radial velocity (V$_\gamma$), the  orbital inclination (i),
the temperature of the secondary component (T$_h$), the potential of the components ($\Omega _{\rm c}$, $\Omega _{\rm h}$),
the monochromatic luminosity of cooler star (L$_{\rm c}$) and spot parameters (latitude, longitude,
size, and temperature factor) were adjustable parameters.
In Table~\ref{tab:lc} we listed the parameters obtained in this study. Last digits show the errors.
Precise and rich data set led us to obtain accurate parameters of the RU Cnc.
\par
\begin{figure}
\centering
\includegraphics[height=60mm]{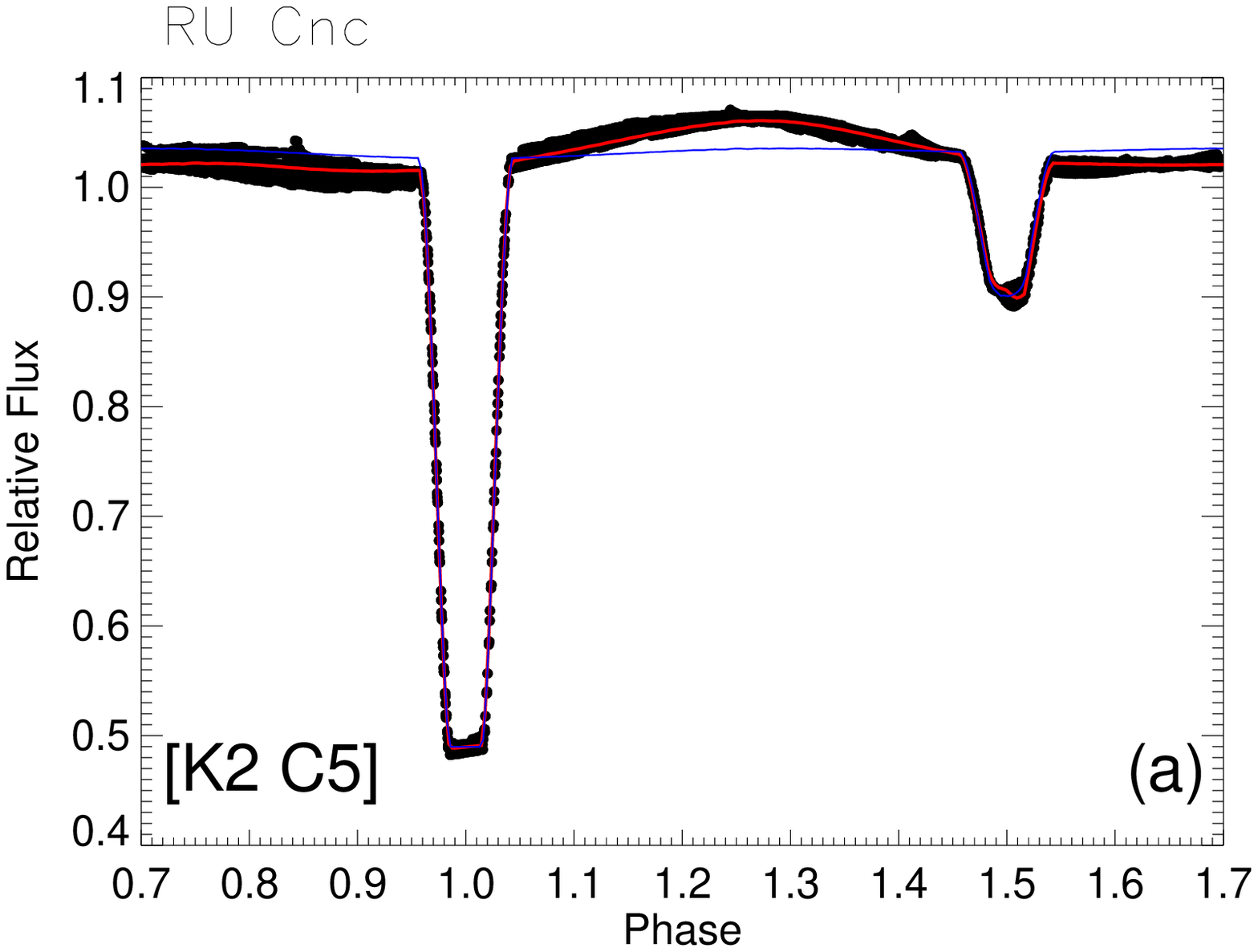}\\
\vspace{-0.3cm}
\includegraphics[height=35mm]{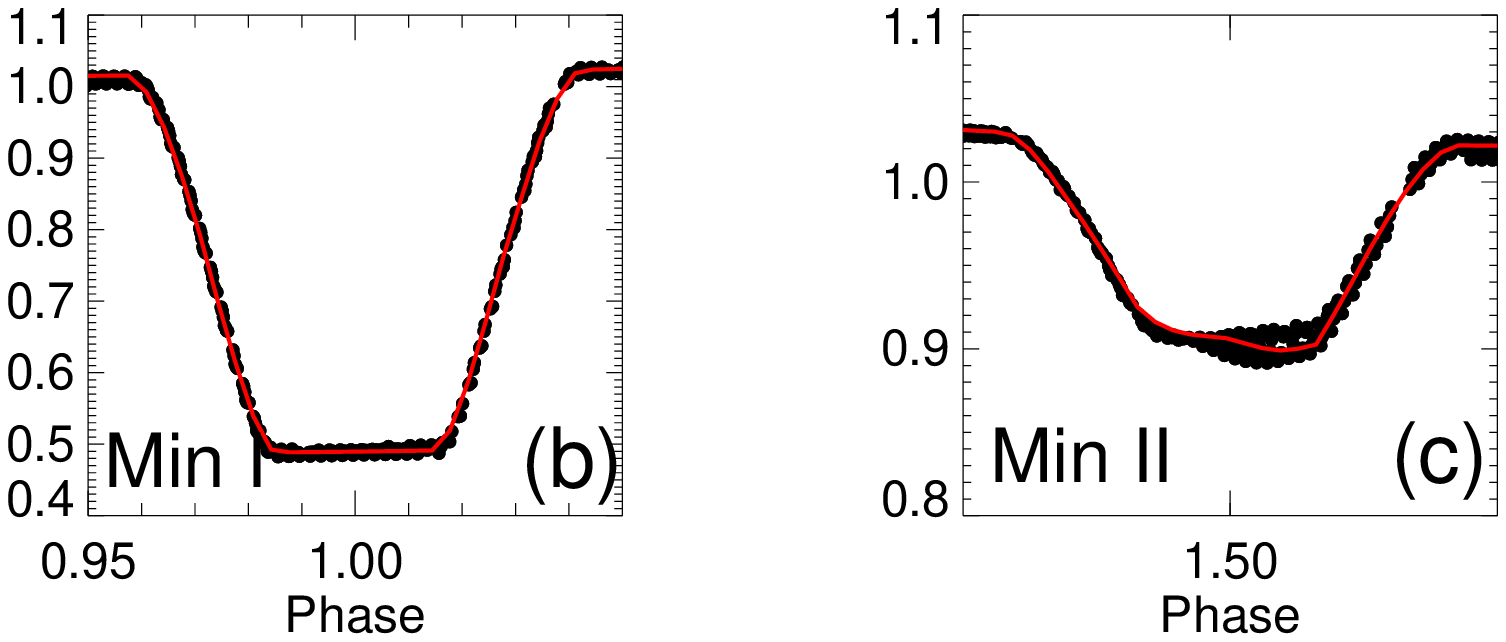}\\
\vspace{-0.3cm}
\includegraphics[height=60mm]{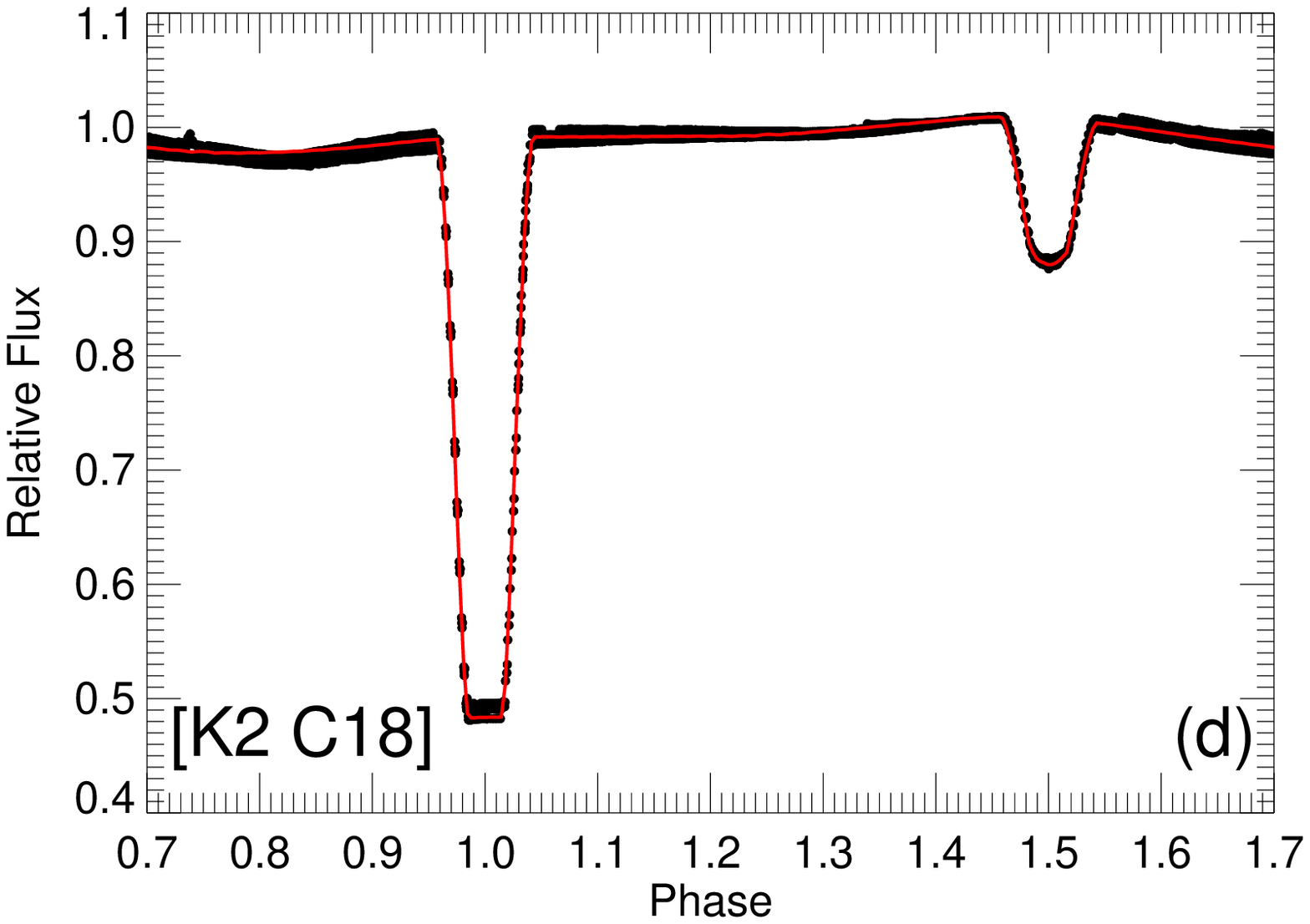}\\
\vspace{-0.3cm}
\includegraphics[height=42mm]{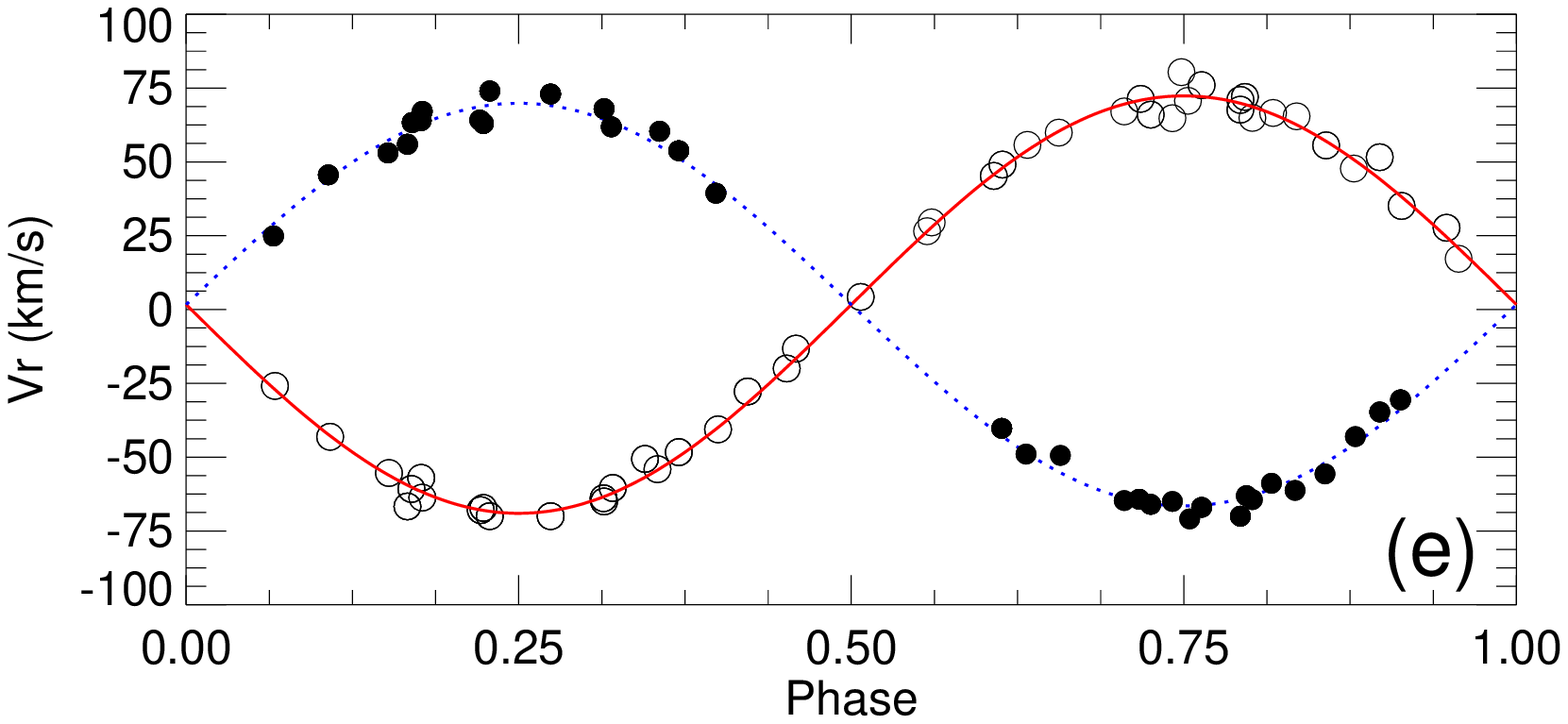}\\
\caption{Observed (dots) and computed (line) light curve of RU Cnc for the K2 C5 (a) and C18 (d).
Primary (b) and secondary (c) minima expanded to emphasize the agreement.
(e) Observed and computed radial velocities taken from Popper (1990) and Imbert (2002).}\label{Fig:rucnc_lcrv}
\end{figure}

The model of the light curve analysis obtained using the values given in Table~\ref{tab:lc} are
shown in Figure~\ref{Fig:rucnc_lcrv}a in comparison with the observations.
Figure~\ref{Fig:rucnc_lcrv}b-c show the eclipses. Both spotted and unspotted models are assumed.
Unspotted model is shown in Figure~\ref{Fig:rucnc_lcrv}a with a blue dashed line and continuous red line show spotted model solution.
Results indicate that it is inevitable to make the assumption of a spotted surface.
\par
Stellar spots are represented usually with  latitute ($\beta$), longitute (($\lambda$), angular radius ($r$) and temperature factor ($t$) parameters (see Pr{\v s}a and Zwitter, 2005; Wilson, 1979) for details. Studies on the binary systems show there is no unique synthetic spot modeling. Although the most probable spot(s) of the observational data can be determined via some testing. The light curve analysis programs do not allow to treat the spot parameters ($\beta$, $\lambda$, $r$, $t$) as free parameters or do not provide reliable results when they are treated as free parameters. On the other hand, Wilson (2012) prosed significant improvement in models with spots. Stellar spots in the Sun and active stars are regions that are generally cooler and sometimes hotter than the rest of the stellar surface with an average life span of a few months. 
\par

The subgiant component of RS Cvn type binary systems has not filled its Roche lobe and show photometric variations (Hall, 1976). 
Observations of these kinds of binary systems with active cold components show chromospheric plages, flares, and coronal activities at different wavelengths.  
Stellar spots in some cases cover a significant percentage of stellar surface (see, Frasca et al., 2008 and references therein). 
In most cases, stellar surfaces contain dark and cool spots; however, in some cases, 
it is also possible to contain structures brighter than the surface called plages. 
These structures are observed in the Sun and other cool stars (see, Berdyugina (2005) and references therein).

In this study, we analyzed the data set of RU Cnc obtained in less than 75 days at three separate times. Following the unspotted model of K2 C5 data, the best solution was obtained, assuming three spots (S1, S2, and S3) on the surface of the cooler component. The spot parameters obtained for the K2 C5 data are summarized in Table \ref{tab:spots}. K2 C5 data set obtained between 27 April 2015 and 15 July 2015. However, K2 C18 data set obtained between 27 May 2018 and 2 July 2018 in 36 days. Therefore, we modelled the K2 C18 data without changing any orbital parameters of the binary system but the spot parameters. 
Our analysis proposes the existence of five relatively small spots on the surface of the secondary star during K2 C18 observations.
Finally, for the orbital and physical parameters obtained from the C5 light curve modeling, we modeled the TUG T60 BVR data, which TUG observations start in the same time with K2 C5 data set. At this point, we would like to underline a known fact about the structures called stellar spots on the stellar surfaces. Instead of representing individually stellar spots it represents a group of stellar spot.

\begin{table}
\caption{Photometric and spectroscopic elements of RU Cnc and their formal
1$\sigma$ errors.  See text for details.}\label{tab:lc}
\begin{tabular}{ll}
\hline
$T_{0}$ as ${\rm JD}/{\rm{d} - 2400000}$   		& 57157.3128(31) \\
$P/{\rm d}$                               		& 10.172918(3)   \\
$i/^\circ$                                	    & 89.7(4)	 \\
$\Omega _{\rm c}$                             & 12.71(27)	  \\
$\Omega _{\rm h}$                             & 6.78(6)	   \\
$q = m_{\rm h}/m_{\rm c}$                       & 1.0366(12)    \\
$T_{\rm h}/{\rm K}$                             & 6860 (285)     \\
$T_{\rm c}/{\rm K}$                             & 4800 (200)   \\
Luminosity ratios:                        		& 				 \\
$l_{\rm c}/l_\textrm{total}$                    & 50.6\%         \\
$l_{\rm h}/l_\textrm{total}$                    & 49.4\%         \\
Fractional radius of primary ($R_{\rm h}/a$)    & 0.0857(20)     \\
Fractional radius of secondary ($R_{\rm c}/a$)  & 0.1797(19)     \\
\hline
$K_{\rm c}/{\rm km\,s^{-1}}$                    & 70.69(8)       \\
$K_{\rm h}/{\rm km\,s^{-1}}$                    & 68.19(8)       \\
System velocity $V_\gamma/{\rm km\,s^{-1}}$  	& $1.66(3)$       \\
a$_{\rm h}$sin$i/R_{_\odot}$                    & 14.208(16)     \\
a$_{\rm c}$sin$i/R_{_\odot}$                    & 13.706(15)     \\
m$_{\rm h}$sin$^3i/M_{_\odot}$                  & 1.386(3)       \\
m$_{\rm c}$sin$^3i/M_{_\odot}$                  & 1.437(3)       \\
\hline
\end{tabular}
\end{table}

\begin{table}
\caption{Spot parameters of the cooler component for K2 C5 and K2 C18 data sets. K2 C5  data is obtained at about 75 days from 2457139.6 to 2457214.4JD. K2 C18 data is obtained after 3 years from the C5 data at 36 days from 58266.0 to 58302.4. See text for details.}\label{tab:spots}
\begin{tabular}{lllll}
\hline
Spot 				& Colatitude 		& Longitude  		& Radius 		& Temperature    	\\
 					& ($^\circ$) 		&($^\circ$)    		& ($^\circ$)   	&(T$_{\textrm{spot}}$/T$_\textrm{c}$)  	\\
\hline
K2 C5  & &	& &	\\
\hline
S1 	&  	80	& 220 	& 15 	& 0.92 	 \\
S2 	&   70  & 175	& 13	& 1.09        \\         	 
S3  &   90	& 280	& 20	& 0.95 		\\                          	   
\hline
K2 C18  & & &	& 	\\
\hline
S4 	&  	90	& 280 	& 23 	& 0.92 	 	\\
S5 	&   75  & 0		& 20	& 1.05       \\         	 
S6  &   90	& 90	& 15	& 0.92 	 		\\
S7 	&   90  & 210	& 10	& 0.94        \\         	 
S8  &   120	& 55	& 10	& 0.93  		\\                             	   
\hline
\end{tabular}
\end{table}

\section{PHYSICAL PARAMETERS OF THE SYSTEM}

The high sensitivity of the {\it{Kepler}} K2 observations for RU Cnc gives us the opportunity
to determine much more accurately the orbital parameters of the binary and the physical parameters of the components.
Radial velocity curves yield the velocity amplitudes of the components as listed in Table~\ref{tab:lc}.

\begin{table}
\scriptsize
\begin{center}
\caption{Absolute parameters of RU Cnc.  Standard errors of 1$\sigma$ in the last digits are given in
parentheses.}\label{tab:PhyPar}
\begin{tabular}{lll}
\hline
                                                  & Hotter Star    		  & Cooler Star     \\
\hline
Masses $M/\rm{M_{_\odot}}$                        & $1.437(46)$     	  & $1.386(44)$     \\
Radii $R/\rm{R_{_\odot}}$                         & $2.392(69)$     	  & $5.016(80)$     \\
Effective temperatures $T_{\rm eff}/{\rm K}$      & $6\,860(285)$    	  & $4\,800(200)$        \\
Luminosities $\log_{10}(L/{\rm L_{_\odot}})$      & $1.056(87)$     	  & $1.079(78)$     \\
Surface gravity $\log_{10}(g/\rm{cm\,s^{-2}})$    & 3.82(35)        	  & 3.20(15)        \\
Absolute bolometric magnitude $M_B$               & 2\fm09(22)      	  & 2\fm03(16)      \\
Absolute visual magnitude $M_V$                   & 2\fm07(21)     		  & 2\fm43(19)      \\
Separation between stars $a/\rm{R_{_\odot}}$      &~~~~~~~~~~~~27.914(16) &			        \\
Distance $d/{\rm pc}$                             &~~~~~~~~~~~~380(57)    &      			\\
\hline
\end{tabular}
\end{center}
\end{table}

Using the newly obtained orbital parameters listed in Table~\ref{tab:lc}
we computed the physical parameters of the components (Table~\ref{tab:PhyPar}).
During our calculations, the temperature of the Sun is taken to be 5772~K and its bolometric magnitude  4.755 mag (Mamajek (2015)).
The bolometric corrections BC$_1$ and BC$_2$ are obtained from Flower (1996) tables corresponding to the temperatures of the components.
Following BC$_1$ and BC$_2$ values estimation we determined the absolute visual magnitudes for
the components and then with the magnitudes and reddening we find the distance of RU Cnc $380\pm57\,$ pc.

\section{DISCUSSION AND CONCLUSION}

Double-lined, eclipsing detached binary systems are the systems whose physical parameters can be determined accurately.
In this sense, RU Cnc is an important system since its physical parameters could be determined very accurately from
simultaneous analysis of two radial velocity curves that were obtained at two separate times with precise Kepler K2 data.
These physical parameters are especially important in contracting stellar evolution models.
The mass and radius of RU Cnc were obtained within \%3 sensitivity;
the results with other physical parameters are given in Table~\ref{tab:PhyPar}. Popper (1990)
obtained the parameters of the components from the radial velocity and not complete light curve analysis.
In this study, the radius of the primary star is estimated to be $8\,$ percent and the secondary star is
estimated to be $2.4\,$ percent larger than in Popper (1990) study. The similar difference appears to be in the masses.
These differences are most probably due to the higher precisely of {\it{Kepler}} K2 data relative to ground-based observations.

The distance of RU Cnc is given as 300 pc (Popper 1990), 331 pc (ESA, 1997), and 279 pc ({van Leeuwen 2007).
In this study, we obtained a distance of $380\pm57$ pc from simultaneous radial and light curve solution.
Recently, Gaia gave a distance of $419\pm6.6$ pc (Gaia Collaboration, 2016) which shows a consistency between the parallax obtained in this study and that of Gaia.
	
The system RU Cnc is a system that has been followed observationally for more than a century.
Eggleton and Yakut (2017) calculated non-conservative evolutionary models of 60 binary systems with giant components.
In the study of Eggleton and Yakut (2017), the authors calculated the age of the system as 2.5 Gyr.
In addition, the authors gave the initial parameters of 11.72 days, 0.3, 1.550, and 1.425 for the orbital period,
eccentricity, the mass of the hot component, and the mass of the cooler component (less massive component in the observations); respectively.
The authors concluded the system consisting of a red giant branch (RGB) and a main-sequence (MS) components. The RGB component, once the massive star, has been lost 15\% of its mass.

Eggleton \& Yakut (2017) discussed a discrepceancy between their model and observation. One of the most important reason for this is can be due to previosly obtained less accurate data. In this study, the physical parameters obtained much more accurately. The radius of the cooler star is estimated to be 3 per cent larger while the hotter star is estimated 8 per cent larger than in their study. The mass of the components is 2 per cent larger and the temperature of the hotter component is 8 per cent larger than their study. These differences are most probably due to the higher accuracy of Kepler data relative to ground-based observations. Hence, new physical parameters are much more in accordance with the models.

\section*{Acknowledgments}
We are very grateful to an anonymous referee for comments and helpful, constructive suggestions which helped us to improve the paper.
This study supported by the Turkish Scientific and Technical Research Council (T\"UB\.ITAK-117F188).
We thank to T\"UB\.ITAK for a partial support in using T60 telescope with project number 16DT60-1114.
KY thanks  the COST Action CA-15117 (CANTATA), CA-16104 (GWVerse) and the Institute of Astronomy Cambridge (IoA) for hospitality and support.
DK, TI, and KAC thanks the  T\"UB\.ITAK- B\.IDEB 2214-A scholarship.


\label{lastpage}
\end{document}